\documentclass[aps,showpacs,11pt]{revtex4}
\usepackage{dcolumn}
\usepackage{graphicx}
\usepackage{amsmath}
\usepackage{amsfonts}
\usepackage{amssymb}
\usepackage{psfrag}
\usepackage{wrapfig}
\usepackage{subfigure}
\usepackage{makeidx}
\usepackage{bm}
\usepackage{epsf}
\usepackage{hyperref}
\usepackage{color}

\begin{document}

\title{Holographic Schwinger effect in a confining background with Gauss-Bonnet corrections}

\author{Shao-Jun Zhang $^{1,2,3}$}
\email{sjzhang84@hotmail.com}
\author{E. Abdalla $^1$}
\email{eabdalla@usp.br}
\affiliation{$^1$ Instituto de
F$\acute{i}$sica, Universidade de S$\tilde{a}$o Paulo, C.P. 66318,
05315-970, S$\tilde{a}$o Paulo, SP, Brazil\\
$^2$ Institute for Advanced Physics and Mathematics, Zhejiang University of Technology, Hangzhou 310032, China\\
$^3$ Kavli Institute for Theoretical Physics China, CAS, Beijing 100190, China
}

\date{\today}

\begin{abstract}
\indent We study the effect of higher-derivative terms on holographic Schwinger effect by introducing the Gauss-Bonnet term in the gravity sector. Anti-de Sitter (AdS) soliton background is considered which is dual to confining phase of the boundary field theory. By calculating the potential between the produced pair, we find that larger Gauss-Bonnet factor $\lambda$ makes the pair lighter. We apply numerical method to calculate the production rate for various cases. The results show that the Gauss-Bonnet term enhances the production rate. The critical behaviors near the two critical values of the electric field are also investigated, and it is found that the two critical indexes are not affected by the Gauss-Bonnet term and thus suggests a possible universality.

\end{abstract}



\maketitle

\section{Introduction}

It is well known that vacuum in quantum field theory (QFT) is not simply empty space. Rather, it contains short-lived virtual pairs of particle and antiparticle due to quantum fluctuations. For example, in the quantum electrodynamics (QED) vacuum, eletron-positron pairs  are momentarily created and annihilated. Moreover, when an external strong electric field is applied, these virtual particles can be materialized and become real particles. This is the well known Schwinger effect~\cite{Heisenberg:1935qt,Schwinger:1951nm}. The production rate per unit time and unit volume $\Gamma$ has been evaluated long time ago in the weak-coupling and weak-field approximation. Later, it is generalized to arbitrary-coupling but weak-field regime and the result is~\cite{Affleck:1981ag}
\begin{eqnarray}\label{ProductionRate}
\Gamma = \frac{(e E)^2}{(2\pi)^3} e^{-\frac{\pi m^2}{e E}+\frac{1}{4} e^2},
\end{eqnarray}
where $e$ and $m$ are the charge and mass of the created particles, respectively. Here $E$ is the external electric field. This novel nonperturbative phenomenon can be explained as a tunneling process, and is not unique to QED but usual for QFTs coupled to an $U(1)$ gauge field.

From the above formula of the production rate, it is easy to see that there exists a critical value of the electric field above which the effect is not exponentially suppressed any more and the vacuum becomes catastrophically unstable. It is interesting to see that there is also a critical electric field (related to the string tension) in string theory, above which open strings with opposite charges residing at the two endpoints respectively become unstable~\cite{Fradkin:1985ys,Bachas:1992bh}. Its analogy with the Schwinger effect in QFTs inspires attempts to realize the latter in the AdS/CFT context~\cite{Maldacena:1997re,Gubser:1998bc,Witten:1998qj}.

The first step to achieve this goal is to realize a system coupled with a $U(1)$ gauge field. In the best established example of AdS/CFT, namely the duality between $N=4$ superconformal Yang-Mill theory (SYM) and type IIB supergravity in $AdS_5 \times S_5$, this can be realized by breaking the gauge group from $SU(N+1)$ to $SU(N) \times U(1)$ via the Higgs mechanism in the CFT side. After the breaking, the fundamental scalar fields in the W-boson supermultiplet, which are often called "W-bosons" or "quarks", are coupled to a $U(1)$ as well as an $SU(N)$ gauge field. In the AdS side, this step amounts to separate one $D3$-brane (the probe $D3$-brane) from the stack of $D3$-branes. Proposed in Ref.~\cite{Semenoff:2011ng} (see also an earlier work~\cite{Gorsky:2001up}), the probe $D3$-brane is placed at a finite radial position rather than at the boundary in order to make the mass of W-bosons finite rather than infinitely heavy. Then the exponential factor in the production rate of W-bosons can be evaluated by calculating the classical Euclidean action of the string worldsheet which is anchored on a circular Wilson loop on the probe $D3$-brane. This calculation has been done for pure AdS background, and then the critical value of the electric field can be read off which agrees with the one obtained by analyzing the Dirac-Born-Infeld (DBI) action. Later, this proposal is generalized to pair-production of other "particles", such as monopole-antimonopole pairs and dyon-antidyon pairs~\cite{Bolognesi:2012gr}, and also to the case with magenetic fields~\cite{Sato:2013pxa}.

This calculation is then generalized from pure AdS to other interesting backgrounds, more specifically backgrounds describing geometries of AdS soliton~\cite{Sato:2013dwa,Kawai:2013xya} (see also~\cite{Ghodrati:2015rta,Wu:2015krf}). These backgrounds, according to the AdS/CFT, are dual to confining phase of QFTs~\cite{Witten:1998zw,Horowitz:1998ha}. The motivation for studying these confining backgrounds comes from QCD, where quarks are usually in a confining phase and the Schwinger effect may provide a new mechanism of a confinement/deconfinement phase transition. It is expected that the Schwinger effect may be observed on RHIC and LHC, where collision of heavy ions always induces strong enough electro-magnetic fields. By applying the prescription described above, the production rate of W-bosons can also be calculated. From the results, it is found that, beyond the usual critical value $E_c$ of the electric field, there exists another threshold $E_s$ below which the Schwinger effect can not occur. This can be seen as a characteristic of confining backgrounds. These two critical values can also be derived by analyzing the behavior of the potential between W-bosons~\cite{Sato:2013iua,Sato:2013hyw}. Moreover, by analyzing the behavior of the classical Euclidean string action near the two critical values, two universal indexes are found which are independent of the location of the probe $D3$-brane. Investigations are also extended to the case where the induced metric on the probe brane is curved~\cite{Fischler:2014ama}, and to Lifshitz and hyperscaling violation geometries~\cite{Fadafan:2015iwa}. For a recent review on this topic, see Ref.~\cite{Kawai:2015mha}.

As the holographic setup of the Schwinger effect is closely related to the string theory, it is natural to take into account various stringy corrections. In this paper, we would like to consider one stringy correction, more specifically, the effect of Gauss-Bonnet terms whose existence will modify the confining backgrounds. On the field theory side, it amounts to consider the effect of finite (but large) 't Hooft coupling. We would like to see how this higher-derivative term affects the properties (the mass for example) of W-bosons and the Schwinger effect.  Moreover, it is also interesting to see whether the two universal critical indexes are modified or not. These are the main motivations of the present work.

The paper is organized as follows. In the next section, we briefly introduce the Gauss-Bonnet soliton background. In Sec. III, by analyzing the potential between W-bosons holographically, the mass of W-bosons and the two critical values of the electric field are obtained. In Sec. IV, we calculate the production rate of W-bosons holographically and analyze the critical behaviors. The last section is devoted to summary and discussions.

\section{Confining background in Gauss-Bonnet gravtiy}

The action of Gauss-Bonnet gravity in $(d+1)$ dimensions $(d\geq 4)$ is
\begin{eqnarray}\label{action}
S = \frac{1}{16 \pi G_N^{(d+1)}} \int d^{d+1}x \sqrt{-g} \left[\frac{d(d-1)}{L^2} +R + \frac{L^2 \lambda}{(d-2)(d-3)} \left(R_{\mu\nu\rho\sigma} R^{\mu\nu\rho\sigma}- 4 R_{\mu\nu} R^{\mu\nu} +R^2\right)\right],
\end{eqnarray}
where the first term is the cosmological constant term, $\Lambda = - \frac{d(d-1)}{2 L^2}$, $\lambda$ is the Gauss-Bonnet factor, which is usually constrained in the range~\cite{Buchel:2009tt,Camanho:2009vw,Buchel:2009sk}
\begin{eqnarray}\label{lambdaconstraint}
-\frac{(d-2)(3d+2)}{4(d+2)^2} \leq \lambda  \leq \frac{(d-2)(d-3)(d^2-d+6)}{4(d^2-3d+6)^2},
\end{eqnarray}
by respecting the causality of the dual field theory on the boundary and preserving the positivity of the energy flux in CFT analysis.

From the action, we can derive the equations of motion
\begin{eqnarray}\label{EOM}
R_{\mu\nu} - \frac{1}{2} R g_{\mu\nu} - \frac{d(d-1)}{2 L^2} g_{\mu\nu} + \frac{L^2 \lambda}{(d-2)(d-3)} H_{\mu\nu}=0,
\end{eqnarray}
where
\begin{eqnarray}
H_{\mu\nu} = 2 \left(R_{\mu\rho\sigma\xi} R_\nu^{~\rho\sigma\xi}-2 R_{\mu\rho\nu\sigma} R^{\rho\sigma} -2 R_{\mu\rho} R^\rho_{~\nu} + R R_{\mu\nu}\right)-\frac{1}{2} \left(R_{\alpha\beta\rho\sigma}R^{\alpha\beta\rho\sigma} - 4 R_{\alpha\beta} R^{\alpha\beta} + R^2\right) g_{\mu\nu}.\nonumber\\
\end{eqnarray}
The above equations of motion admit an AdS soliton solution~\cite{Cai:2007wz}
\begin{eqnarray}\label{soliton}
ds^2 &=& \frac{L^2}{z^2} \left[f(z) d\phi^2 + f^{-1}(z) dz^2 + L^2 \left(-dt^2 + \delta_{ij} dx^i dx^j \right)\right],\\
f(z) &=& \frac{1}{2\lambda} \left(1-\sqrt{1+ 4 \lambda \left(\frac{z^d}{z_s^d}-1\right)}\right),\nonumber
\end{eqnarray}
which can be obtained from the Gauss-Bonnet black hole solution~\cite{Cai:2001dz} by a double Wick rotation. The geometry terminates at $z=z_s$. Requiring regularity of the geometry at this point will make the coordinate $\phi$ to be periodic, $\phi \sim \phi + \beta_s$ with $\beta_s = \frac{4\pi}{d} z_s$. According to AdS/CFT, this geometry is dual to the confining phase of field theory~\cite{Witten:1998zw,Horowitz:1998ha}.

In the following sections, we will only consider the case with $d=4$, that is the dual field theory we consider is $(3+1)$-dimensional. Then, from Eq.~(\ref{lambdaconstraint}), the Gauss-Bonnet factor is constrained in the range $-\frac{7}{36} \leq \lambda \leq \frac{9}{100}$. We choose the convention that $L=1$. We work with Euclidean signature which can be achieved by the Wick rotation $t \rightarrow -i\tau$.

\section{Potential analysis}

In this section, we would like to analyze the potential between produced W-bosons holographically. We consider a rectangular Wilson loop on the probe brane located at $z=z_0$. Without loss of generality, we can put it on the plane spanned by $\tau-x_1$:
\begin{eqnarray}
{\cal C}: \quad 0 \leq \tau \leq T,\quad -\frac{x}{2}\leq x_1\leq \frac{x}{2}
\end{eqnarray}
where $x_1$ is one of the un-compacted coordinates on the brane. $T$ and $x$ denote the length and width of the Wilson loop respectively, and $T\gg x$. It is well known that the expectation value of the Wilson loop is related to the potential between W-bosons $V$ (including the static energy of the pair) as $\langle W({\cal C}) \rangle \sim e^{-T V}$. On the other hand, according to the AdS/CFT duality, we have $\langle W({\cal C}) \rangle \sim e^{-S_{NG}}$ in the classical limit~\cite{Rey:1998ik,Maldacena:1998im}, where $S_{NG}$ is the classical string action with the worldsheet anchored on the Wilson loop. So we can evaluate the potential by calculating the classical string action. Considering the symmetry of the background, the string worldsheet can be parameterized by only one function, $z=z(x_1)$, and the induced metric on the world sheet is
\begin{eqnarray}
ds^2 = \frac{1}{z^2} \left[d\tau^2 + \left(1+\frac{z'^2}{f(z)}\right) dx_1^2\right],
\end{eqnarray}
where prime means derivative with respect to $x_1$, i.e., $'\equiv \frac{d}{dx_1}$. Then the classical string action is
\begin{eqnarray}
S_{NG} = T_F \int d\tau d x_1 \sqrt{{\rm det} G_{ab}} = 2 T_F T \int_0^{x/2} dx_1 \frac{1}{z^2} \sqrt{1+\frac{z'^2}{f(z)}},
\end{eqnarray}
where $T_F$ is the string tension and ${\rm det} G_{ab}$ is the determinant of the induced metric on the worldsheet. The classical string configuration $z=z(x_1)$ can be derived by minimizing the action. Note that the integrand in the action does not depend on $x_1$ implicitly, so there is a conservation equation
\begin{eqnarray}
1+\frac{z'^2}{f(z)} = \left(\frac{z_c}{z}\right)^4,
\end{eqnarray}
with $z_c$ being the tip of the string configuration $z_c \equiv z(0)$. Using the above conservation equation, we have
\begin{eqnarray}
\frac{dz}{dx_1} = - \sqrt{f(z) \left[\left(\frac{z_c}{z}\right)^4 -1\right]},
\end{eqnarray}
from it the distance between W-bosons can be expressed as
\begin{eqnarray}
x = 2 a z_0 \int^1_{1/a} \frac{dy}{\sqrt{f(y) (y^{-4}-1)}},
\end{eqnarray}
where the following dimensionless parameters have been introduced,
\begin{eqnarray}
y \equiv \frac{z}{z_c},\quad a\equiv \frac{z_c}{z_0},\quad b\equiv \frac{z_s}{z_0},
\end{eqnarray}
and $f(y)=\frac{1}{2\lambda}\left(1-\sqrt{1+4\lambda\left[\left(\frac{a}{b}\right)^4 y^4 -1\right]}\right)$. From above definitions, it is easy to see that $1 \geq y \geq \frac{1}{a}$ and $b\geq a > 1$. Then the potential, including the static energy of the pair, is
\begin{eqnarray}
V = \frac{2 T_F}{a z_0} \int^1_{1/a} \frac{dy}{y^2 \sqrt{f(y) (1-y^4)}}.
\end{eqnarray}
In the limit of $a \rightarrow b$ (which corresponds to $x \rightarrow \infty$), the potential up to order ${\cal O} (\lambda^2)$ is given by
\begin{eqnarray}
V= \frac{T_F}{z_s^2} x + \frac{T_F}{z_s} \left[2(b-1)-\left(b+\frac{1}{3 b^3}-\frac{4}{3}\right) \lambda\right] + {\cal O}(\lambda^2).
\end{eqnarray}
The first term represents a linear potential characterizing the confining phase. Up to order ${\cal O} (\lambda^2)$, the second term can be explained as the static energy of the pair,
\begin{eqnarray}\label{mass}
2 m_W \equiv \frac{T_F}{z_s} \left[2(b-1)-\left(b+\frac{1}{3 b^3}-\frac{4}{3}\right) \lambda\right].
\end{eqnarray}
Note that the coefficient of the $\lambda$-term in the square bracket, $-\left(b+\frac{1}{3 b^3}-\frac{4}{3}\right)$, is negative which means increasing $\lambda$ will decrease the mass of the pair. From the above equation, we can see that the mass of the pair is related to the location of the probe brane $z_0$ via the parameter $b$ when $z_s$ is fixed. When we are moving the probe brane towards to the boundary, $m_W$ becomes heavier and heavier and eventually are infinite.

By including the effect of an external electric field $E$ along the $x_1$ direction (Later, we will see how this external electric field appears naturally), the total potential is
\begin{eqnarray}
V_{tot}=\frac{2 T_F}{a z_0} \int^1_{1/a} \frac{dy}{y^2 \sqrt{f(y) (1-y^4)}}-\frac{2 T_F \alpha a}{z_0}  \int^1_{1/a} \frac{dy}{\sqrt{f(y) (y^{-4}-1)}}.
\end{eqnarray}
where $\alpha \equiv \frac{E}{E_c}$ with $E_c = \frac{T_F}{z_0^2}$ being the critical electric field above which the vacuum is catastrophically unstable.

According to the analysis of the potential for general backgrounds in Ref.~\cite{Sato:2013hyw}, there are two critical values of the electric field, $E_s = \frac{T_F}{z_s^2}$ and $E_c = \frac{T_F}{z_0^2}$. When $E<E_s$, the pair is confined and no Schwinger effect can occur. When $E_c>E>E_s$, there is Schwinger effect but is exponentially suppressed, While $E>E_c$ the vacuum is catastrophically unstable. In Fig. 1, we plot the total potential for some chosen parameters. The Schwinger effect can be understood as a tunneling process in the presence of the potential. From the figure, we can indeed see that when $\alpha<0.25 (E<E_s)$, the potential is divergent at infinity and the produced pair has no chance to escape to infinity. When $\alpha=0.25 (E=E_s)$, the potential becomes flat as $x\rightarrow \infty$. When $1.0>\alpha>0.25 (E_c>E>E_s)$, there is the potential barrier and the Schwinger effect can occur as tunneling process. While $\alpha>1.0 (E>E_c)$, the barrier vanishes and the vacuum becomes unstable catastrophically.

\begin{figure}
\includegraphics[width=0.4\textwidth]{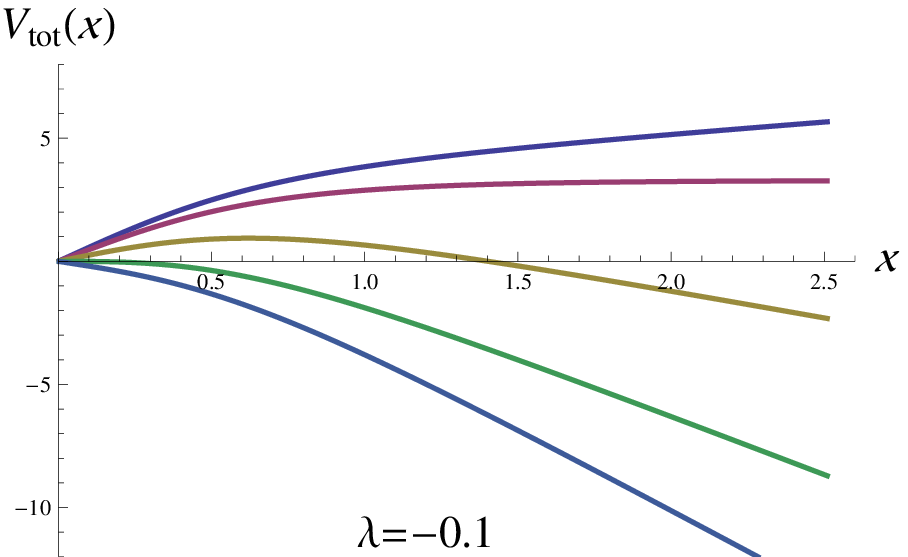}
\includegraphics[width=0.53\textwidth]{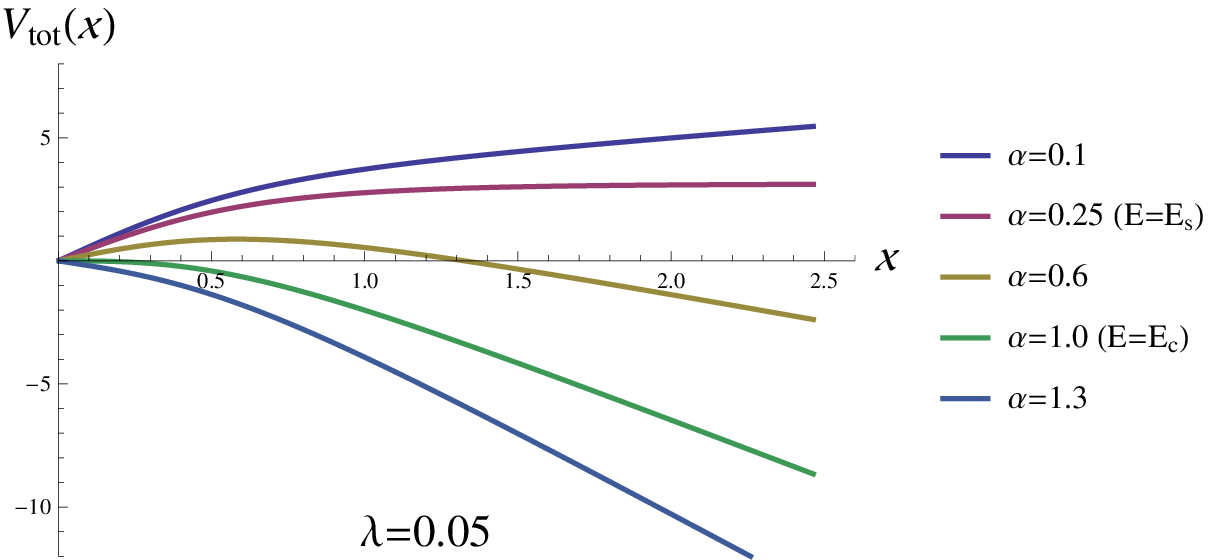}
\caption{Total potential as a function of the distance between the pair. We choose $2\pi T_F =10, z_0=0.5$ and $b=2$.}
\end{figure}

\section{Holographic Schwinger effect}

\subsection{Production rate}

According to the prescription in Ref.~\cite{Semenoff:2011ng}, the Schwinger effect can be evaluated by calculating the expectation value of a circular Wilson loop on the probe brane. As in the previous section, we choose the circular Wilson loop to lie in the $\tau-x_1$ plane. For convenience, we introduce polar coordinates $(\rho, \theta)$ in the plane. And then the string worldsheet can be parameterized by only one function: $z=z(\rho)$. The induced metric on the worldsheet is
\begin{eqnarray}
ds^2 = \frac{1}{z^2} \left[\left(1+\frac{z'^2}{f(z)}\right) d\rho^2 + \rho^2 d\theta^2\right].
\end{eqnarray}
The holographic pair production rate (per unit time and unit volume) is
\begin{eqnarray}
&\Gamma \sim e^{-(S_{NG}+ S_{B_2})},\nonumber&\\
&S_{NG} = T_F \int \sqrt{{\rm det}G_{ab}},\quad S_{B_2} = -T_F \int B_2,&
\end{eqnarray}
where the integral is done over the string worldsheet. The string action contains two parts. The first part is the usual Nambu-Goto action, and the second part represents the coupling with the NS-NS (NS stands for Neveu-Schwarz) $2$-form, $B_2 = B_{01} d\tau \wedge dx^1$, which on the probe brane plays the role of the external electric field. With the induced metric, the two parts can be expressed as
\begin{eqnarray}
S_{NG} &=& 2\pi T_F \int_0^x d\rho \frac{\rho}{z^2} \sqrt{1+\frac{z'^2}{f(z)}},\\
S_{B_2} &=& -2\pi T_F B_{01} \int_0^x d\rho\ \rho = -\pi E x^2,
\end{eqnarray}
where we define the external electric field as $E \equiv T_F B_{01}$. Here, we use $x$ to denote the radius of the circular Wilson loop. The classical string configuration $z=z(\rho)$ can be derived by minimizing the action which gives the equation of motion as
\begin{eqnarray}
\rho z''+ z' + \frac{2 \rho z'^2}{z}-\frac{\rho z'^2}{2 f(z)}\frac{df(z)}{dz} + \frac{2 \rho f(z)}{z}=0.
\end{eqnarray}
To solve this equation, suitable boundary conditions are needed. Note that the equation of motion is singular at $\rho=0$, so to make our numerical calculation available we impose the boundary conditions at the neighborhood of $\rho=0$,
\begin{eqnarray}
z(\epsilon)=z_c + {\cal O} (\epsilon^2), \quad z'(\epsilon)={\cal O}(\epsilon),
\end{eqnarray}
where $\epsilon$ is a small parameter with order of $10^{-3}$ typically. The higher order terms can be derived by solving the equation of motion around $\rho=0$ order by order. $z_c=z(0)$ is the tip of the worldsheet and determined by the constraint
\begin{eqnarray}
z(\rho=x)=z_0.
\end{eqnarray}

After solving the equation of motion with the boundary conditions, we get the classical string configuration $z = z(\rho)$ and then the whole action, which will be a function of the radius of the circular Wilson loop $x$. And then according to the prescription in Ref.~\cite{Semenoff:2011ng}, one further step is needed which is to choose the value of $x$ minimizing the action. This step would make the numerical analysis much harder. However, this step can be replaced by imposing an additional condition to the classical string solution as pointed out in Ref.~\cite{Sato:2013pxa}. This is the mixed boundary condition for the string coordinates in the presence of $B_2$ and leads to the constraint condition (This mixed boundary condition is applicable for general backgrounds with metric taking the form as Eq.~(\ref{soliton})),
\begin{eqnarray}
z' (\rho=x)= - \sqrt{f(z) \left(\frac{1}{\alpha^2}-1\right)} \ \Bigg|_{z=z_0}.
\end{eqnarray}

To do numerical calculations and compare with the Einstein case in Ref.~\cite{Kawai:2013xya}, we set the endpoint of the soliton as $z_s=1$ and the string tension $2\pi T_F =10$. Then the pair production rate depends on the remaining three parameters: $\alpha, \lambda$ and $z_0$. We calculate numerically the classical action $S$ and the exponential factor $e^{-S}$ for various $\alpha, \lambda$ and $z_0$. In Fig.~2, we give an sample of various results where we plot $S$ and $e^{-S}$ as a function of $\alpha$ with fixed $\lambda=0.05$. Other cases with different $\lambda$ have similar picture. From the figure, we can see that the action $S$ grows up quickly when $\alpha<1.0$ and diverges when $\alpha$ is below a certain value thus leads to a vanishing $e^{-S}$. From analysis in previous section, we know that when $\alpha<\alpha_s$, with $\alpha_s \equiv \frac{E_s}{E_c}=\left(\frac{z_0}{z_s}\right)^2$, there is no Schwinger effect. Our numerical results agree well with the analysis.

\begin{figure}[!htbp]
\includegraphics[width=0.4\textwidth]{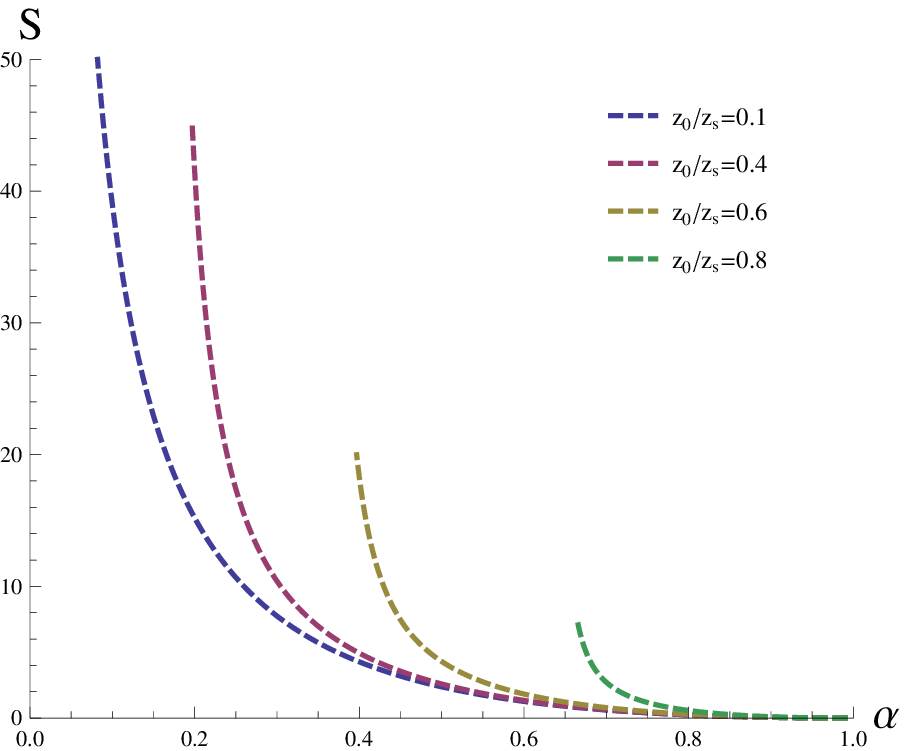}\quad
\includegraphics[width=0.42\textwidth]{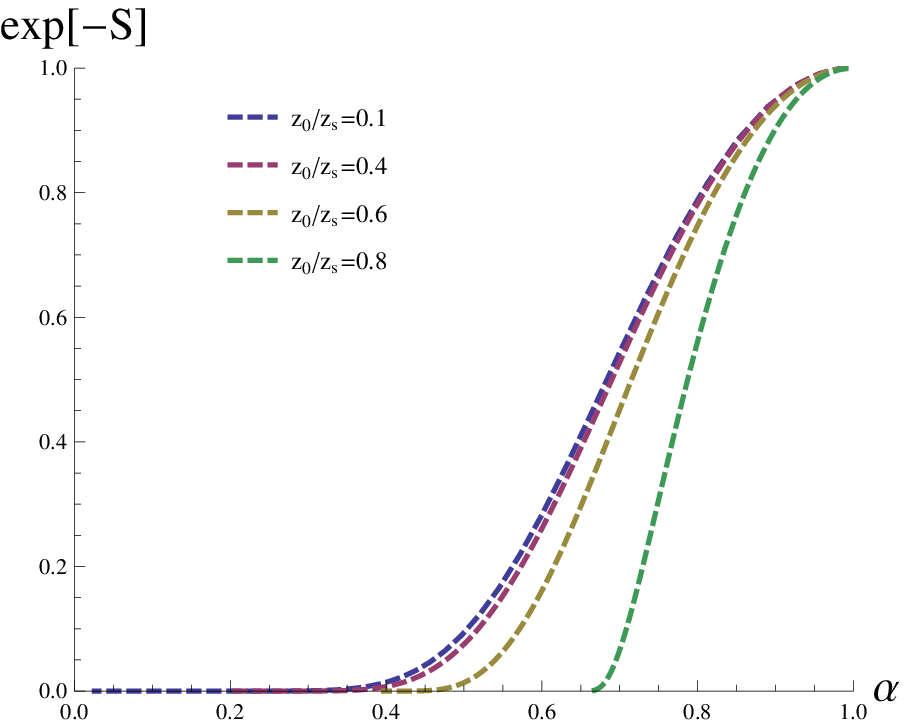}
\caption{The classical action $S$ and the exponential factor $e^{-S}$ as a function of $\alpha$. The Gauss-Bonnet factor is fixed to be $\lambda=0.05$.}
\end{figure}

\begin{figure}[!htbp]
\includegraphics[width=0.4\textwidth]{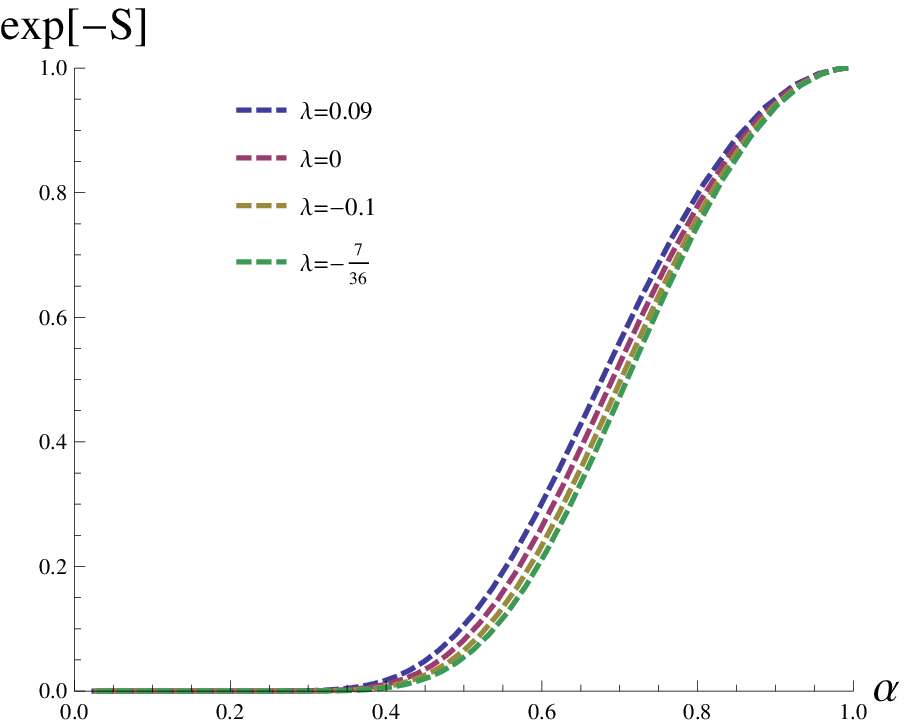}\qquad
\includegraphics[width=0.4\textwidth]{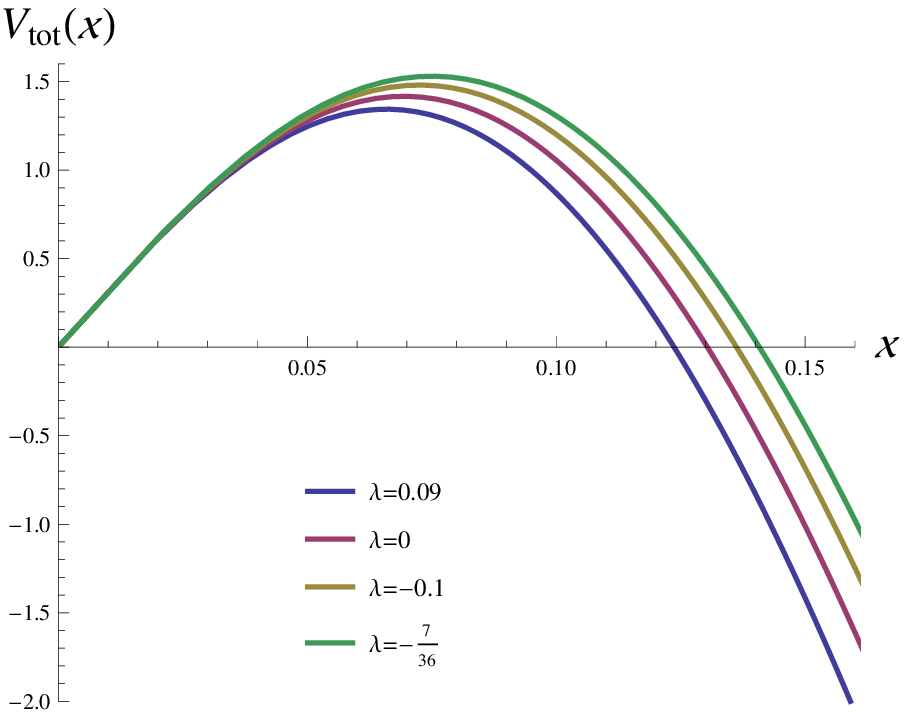}
\caption{{\em Left:} The exponential factor $e^{-S}$ as a function of $\alpha$ for various $\lambda$. {\em Right:} Total potential as a function of the distance between the quark and antiquark. $\alpha$ is chosen to be $0.8$. In both panels, $z_0/z_s$ is fixed to be $0.1$.}
\end{figure}

To see the effect of the Gauss-Bonnet term, in Fig.~3 we fix $z_0/z_s =0.1$ and plot the exponential factor $e^{-S}$ as a function of $\alpha$ for chosen various $\lambda$. From the figure, we can see that for fixed $\alpha$, larger $\lambda$ makes $e^{-S}$ larger. This means that the existence of the Gauss-Bonnet term enhances the Schwinger effect. This can be understood by looking at the total potential, as shown in the right panel of Fig.~3, from which we can see that as $\lambda$ increases the height and width of the potential barrier both decreases thus making the produced pair easier to escape to infinity. This can also be understood from Eq.~(\ref{mass}), where we mention that larger $\lambda$ makes the mass of the pair smaller, and thus it is easier to create them from vacuum. As the Gauss-Bonnet term can be viewed as a stringy effect ($\alpha'$-corrections, more precisely), it means that the stringy effect strengthens the Schwinger effect.

Moreover, as we moving the probe brane towards to the endpoint of the soliton, $z_0/z_s$ increases and the effect of the Gauss-Bonnet term becomes weaker. We can see this phenomenon in Fig.~4, where $z_0/z_s=0.8$. From the left panel, we can see that the four curves with different $\lambda$ are nearly overlapped with others. We can also understand this by looking at the total potential, as shown in the right panel. Compared to Fig.~3, we can see that the difference between the height of the potential barrier with different $\lambda$ becomes smaller as we increase $z_0/z_s$.

\begin{figure}[!htbp]
\includegraphics[width=0.4\textwidth]{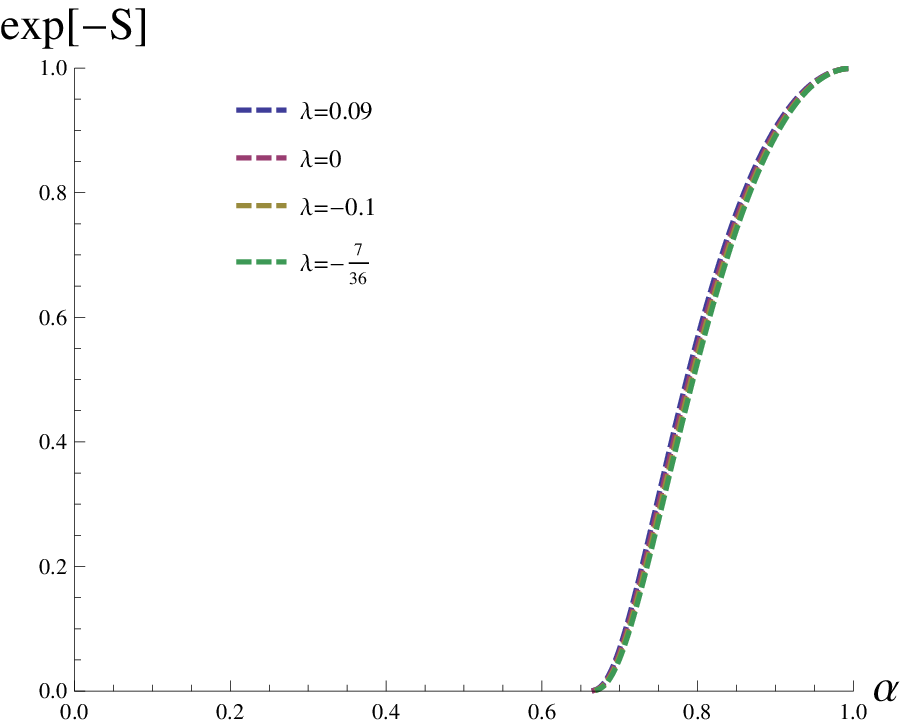}\qquad
\includegraphics[width=0.4\textwidth]{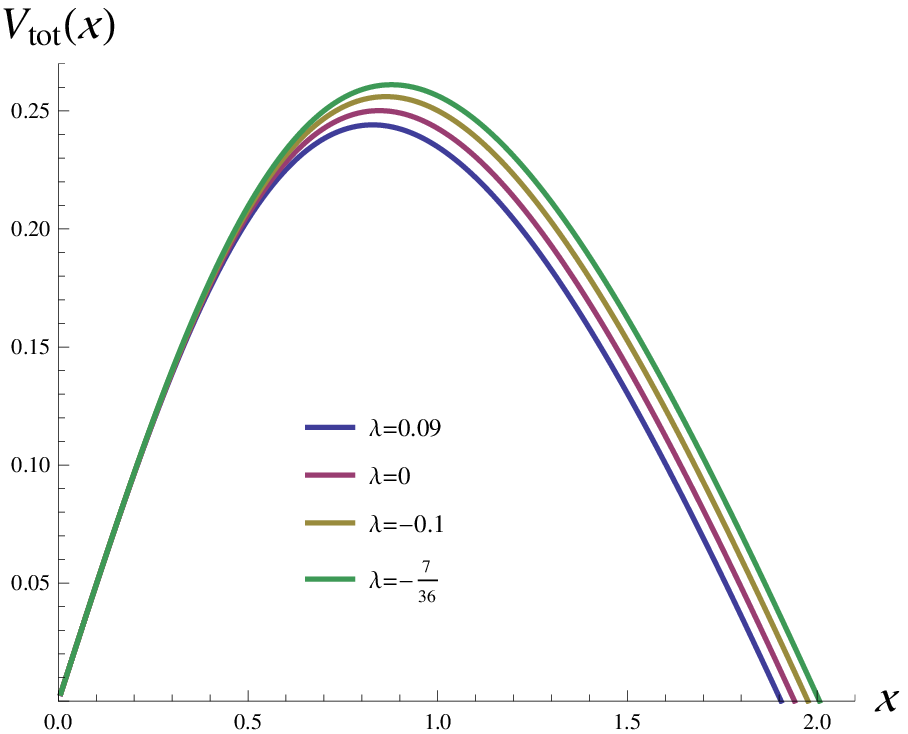}
\caption{{\em Left:} The exponential factor $e^{-S}$ as a function of $\alpha$ for various $\lambda$. {\em Right:} Total potential as a function of the distance between the quark and antiquark. $\alpha$ is chosen to be $0.8$. In both panels, $z_0/z_s$ is fixed to be $0.8$.}
\end{figure}

\subsection{Critical behaviors near $E=E_s$ and $E=E_c$}

In this subsection, we would like to analyze the critical behaviors of the action $S$ near the two critical values of the electric field. From the potential analysis in previous section, we know that the Schwinger effect ceases to disappear when $E$ approaches $E_s$, meaning that $S$ should diverge at $E=E_s$ as
\begin{eqnarray}
S = \frac{C(\alpha_s,\lambda)}{(\alpha - \alpha_s)^{\gamma_s}}+ \rm{subleading\ terms},
\end{eqnarray}
with $\gamma_s$ being a positive index. It is found that, for various $\lambda$, our numerical data near $\alpha=\alpha_s$ can be well fitted by the function $\frac{C(\alpha_s,\lambda)}{(\alpha - \alpha_s)^2}+\frac{D(\alpha_s,\lambda)}{(\alpha - \alpha_s)}$. We show two samples in Fig.~5 and 6 with fixed $\lambda=0.05, -0.1$, respectively.

\begin{figure}[!htbp]
\centering
\subfigure[~~the leading term]{\includegraphics[width=0.42\textwidth]{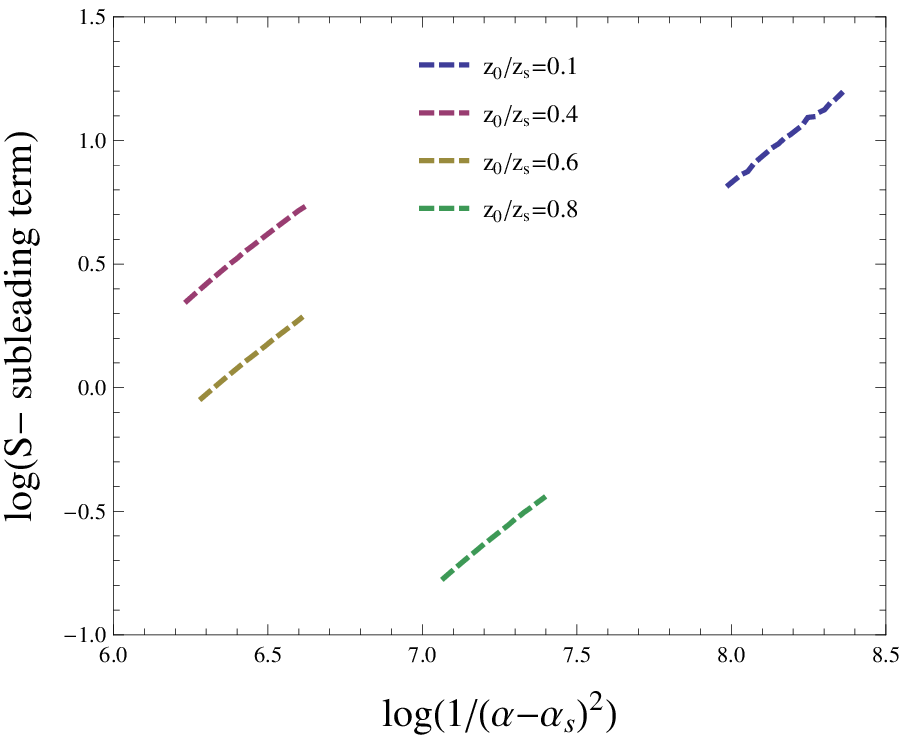}}\qquad
\subfigure[~~the subleading term]{\includegraphics[width=0.4\textwidth]{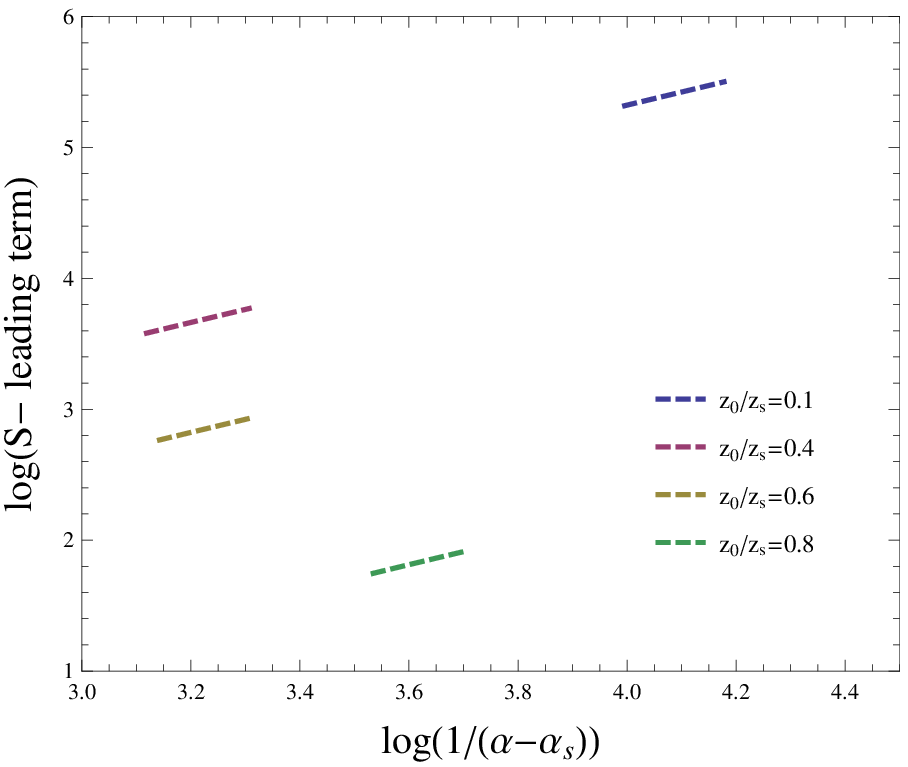}}
\caption{Critical behavior of $S$ near $E=E_s$ with fixed $\lambda=0.05$.}
\end{figure}

\begin{figure}[!htbp]
\centering
\subfigure[~~the leading term]{\includegraphics[width=0.41\textwidth]{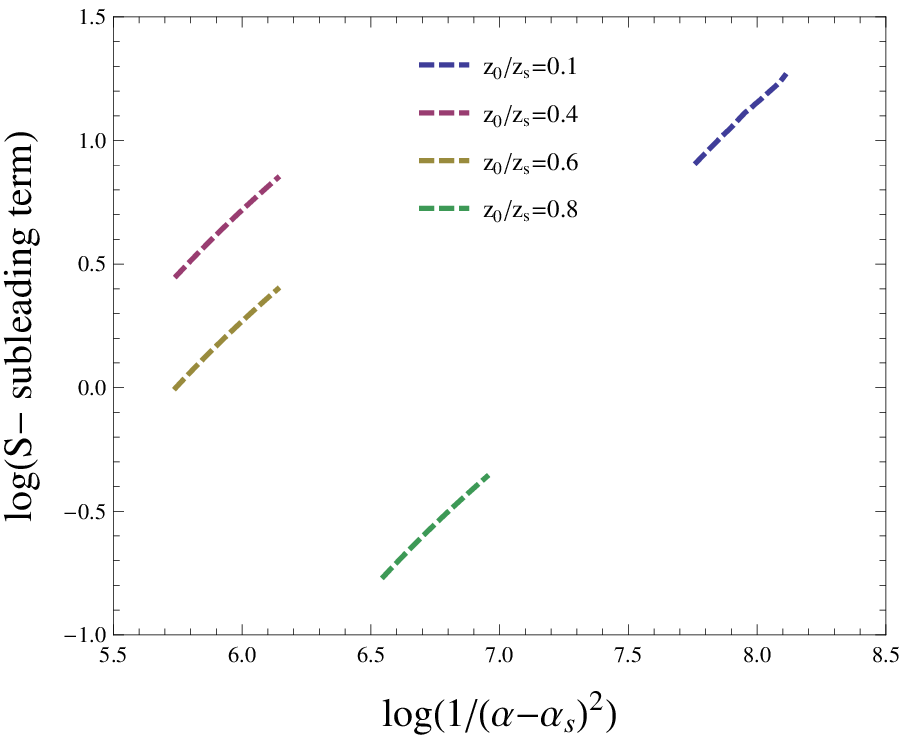}}\qquad
\subfigure[~~the subleading term]{\includegraphics[width=0.4\textwidth]{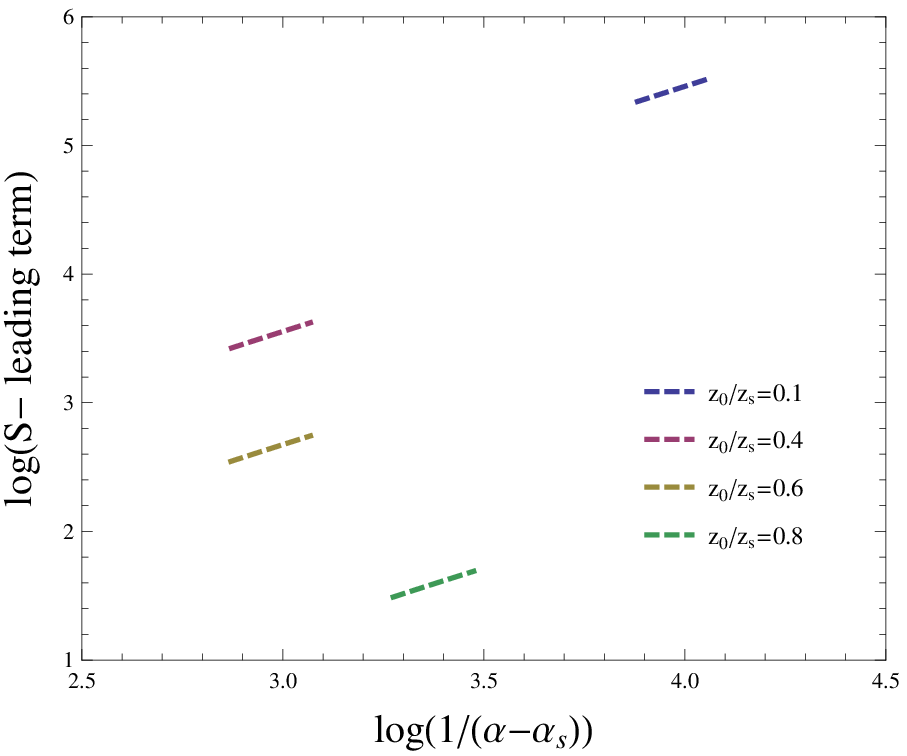}}
\caption{Critical behavior of $S$ near $E=E_s$ with fixed $\lambda=-0.1$.}
\end{figure}

From the figures, we can see that coefficients $C(\alpha_s,\lambda)$ and $D(\alpha_s,\lambda)$ both depend on $\alpha_s$. Also they are found to depend on $\lambda$, as can be read off from Fig.~7 where we plot the fitting results for various $\lambda$. These results suggest a possible universality of the index of the leading term
\begin{eqnarray}
\gamma_s = 2,
\end{eqnarray}
which is the same as in Einstein case Ref.~\cite{Kawai:2013xya} and not affected by the Gauss-Bonnet term. More interesting, it is found that the fitting coefficient $C(\alpha_s, \lambda)$ is very close to zero that has been noticed in Ref.~\cite{Kawai:2013xya} for Einstein case and may suggest the existence of a phase transition.

\begin{figure}[!htbp]
\centering
\subfigure[~~the leading term]{\includegraphics[width=0.4\textwidth]{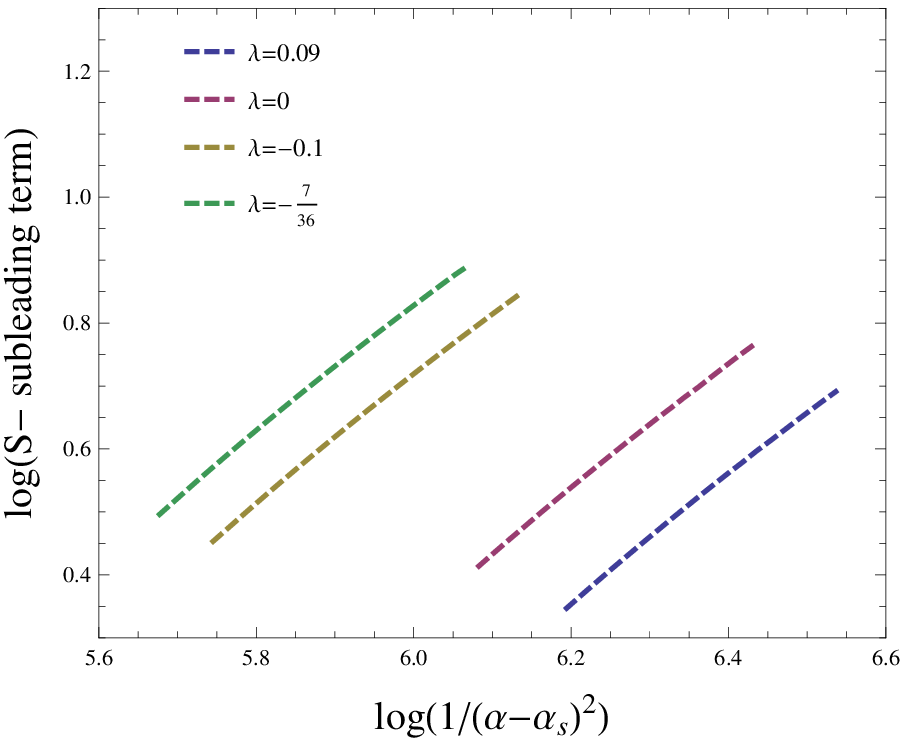}}\qquad
\subfigure[~~the subleading term]{\includegraphics[width=0.4\textwidth]{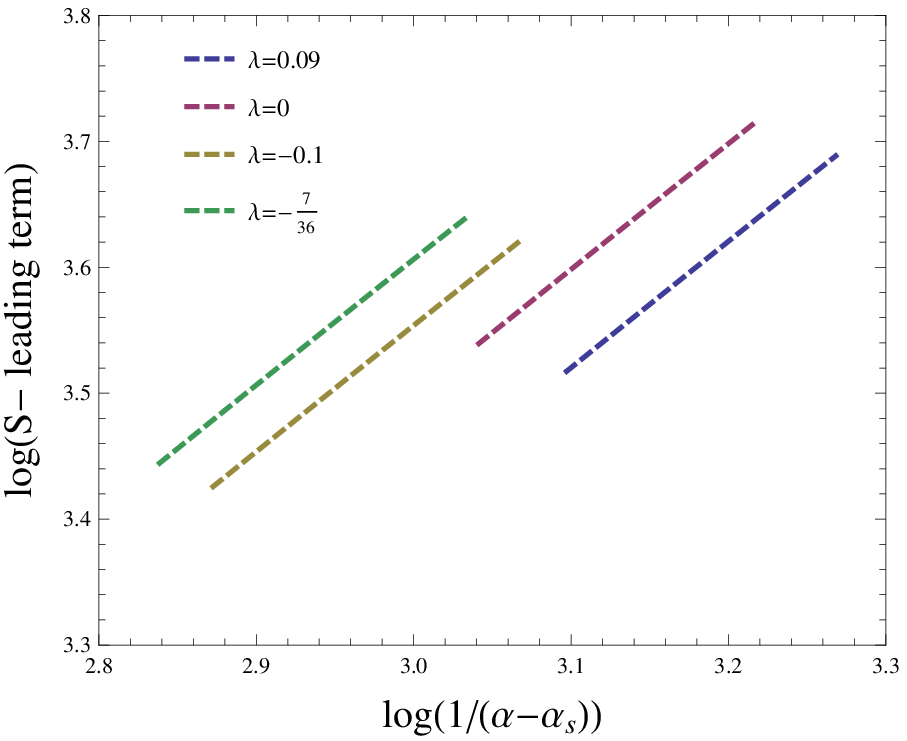}}
\caption{Critical behavior of $S$ near $E=E_s$ with fixed $\alpha_s=0.16 (z_0/z_s=0.4)$.}
\end{figure}

Next, let us discuss the critical behavior of $S$ near $E=E_c$. As stated above, the background becomes catastrophically unstable when $E>E_c$, which means $S$ has the following behavior near $\alpha=1$,
\begin{eqnarray}
S = B(\alpha_s, \lambda) (1-\alpha)^{\gamma_c} + \cdots.
\end{eqnarray}
It is found that our numerical data near $\alpha=1$ for various cases can be well fitted by the function $B(\alpha_s, \lambda) (1-\alpha)^2$, as shown in the Fig.~8 where we give two samples with $\lambda=0.05, -0.1$ respectively. These results suggest a possible universality of the index
\begin{eqnarray}
\gamma_c =2,
\end{eqnarray}
which is also the same as in Einstein case Ref.~\cite{Kawai:2013xya} and not affected by the Gauss-Bonnet term. It is also found that the coefficient $B(\alpha_s, \lambda)$ not only depends on $\alpha_s$ but also on $\lambda$, as can be seen from the Fig.~9.

\begin{figure}[!htbp]
\centering
\includegraphics[width=0.4\textwidth]{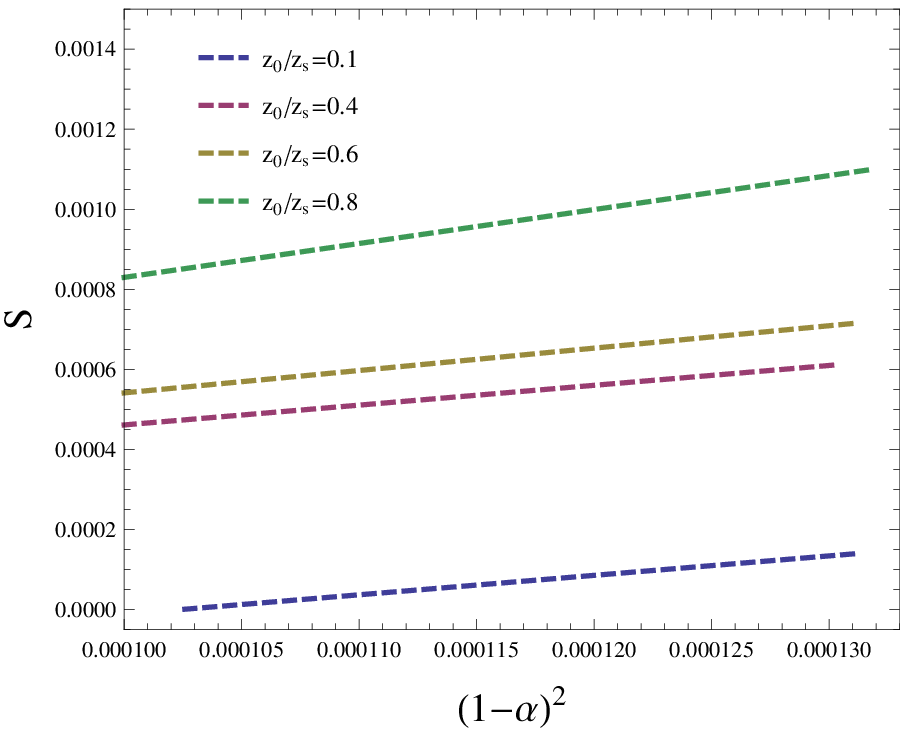}\quad
\includegraphics[width=0.4\textwidth]{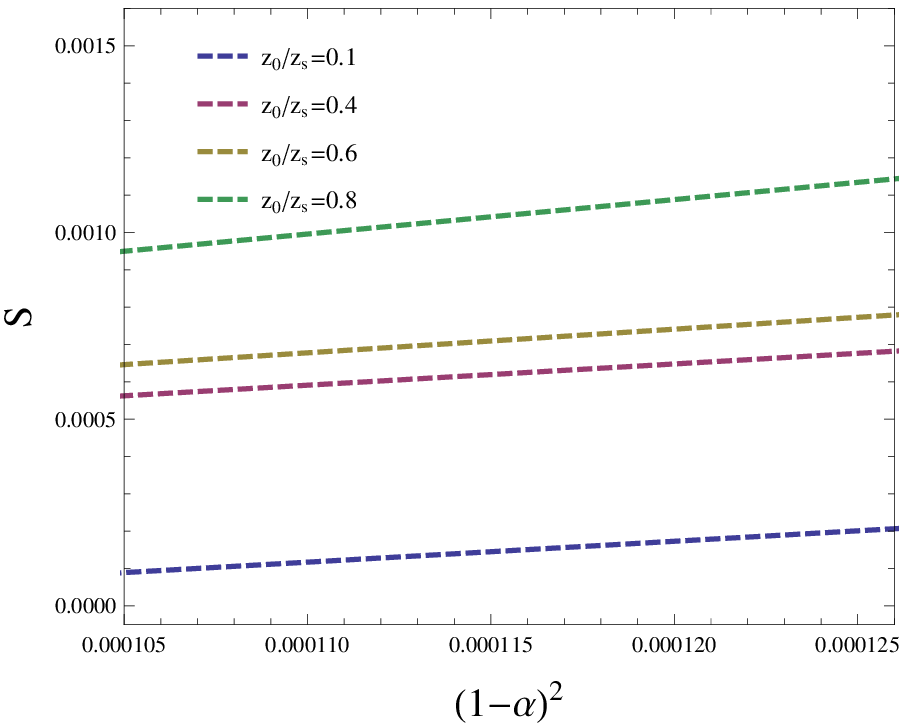}
\caption{Critical behavior of $S$ near $E=E_c$ with fixed $\lambda=0.05$ (left) and $\lambda=-0.1$ (right), respectively.}
\end{figure}

\begin{figure}[!htbp]
\centering
\includegraphics[width=0.5\textwidth]{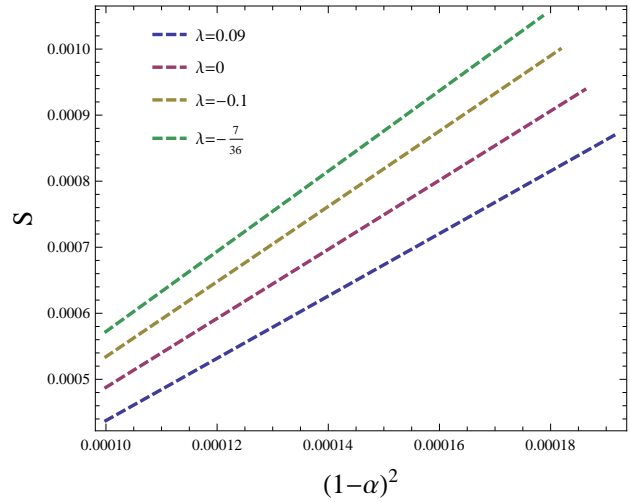}
\caption{Critical behavior of $S$ near $E=E_c$ with fixed $\alpha_s=0.16 (z_0/z_s=0.4)$.}
\end{figure}

\section{Summary and conculsions}

In this paper, we consider the effect of higher-derivative terms on holographic Schwinger effect by introducing Gauss-Bonnet term in the gravity sector. We consider a soliton background which is dual to confining phase of the field theory. Along with the prescription in Ref.~\cite{Semenoff:2011ng}, we calculate the potential between the produced pair holographically, from which we can read off the mass of the pair. We note that larger $\lambda$ makes the pair lighter. Moreover, the two critical values of the electric field can also be obtained from the potential, with one of which characterizing the confining phase. Then we holographically calculate the production rate by using numerical method. From the results, we can see that the Gauss-Bonnet term indeed affect the production rate. Larger $\lambda$ makes $e^{-S}$ bigger, which means the existence of the Gauss-Bonnet term enhances the Schwinger effect. By fitting our numerical data near the two critical values of the electric field, we can obtain the two critical indexes. It is interesting to find that they both are not affected by the Gauss-Bonnet term, which suggests a possible universality. How to interpret their universality is still unknown. It is interesting to see if this universality still holds in other backgrounds, such as dilaton deformed AdS solition~\cite{Cai:2007wz}. This can be left as a further investigation.

It is also interesting to note that the dynamics of confinement is more complex than found by tree-level arguments. This has been reviewed long time ago in lower dimensional models~\cite{Lowenstein:1971fc,Abdalla:2001} and in the first idea about quark confinement~\cite{Kogut:1974sn} as well as in the question of confinement versus screening problem~\cite{Abdalla:2001,Rothe:1978fi,Abdalla:1996ks,Abdalla:1997jp}, which relies deeply on the dynamics of the quantum fields. It is fortunate that the AdS/CFT technique provides means to handle these questions with classical methods.

\section*{Acknowledgement}

This work has been supported by CNPq (Brazil). SJZ Thanks the warm hospitality of Kavli Institute for Theoretical Physics China (KITPC) where part of the work was performed.

\end{document}